\documentclass[aps,prl,reprint,groupedaddress]{revtex4-1}
\usepackage{amsmath,amssymb,graphics,verbatim,epsfig}

\begin{document}

%\preprint{}

\title{Strong bounds on Onsager coefficients and efficiency\\for three terminal thermoelectric transport in a magnetic field}


\author{Kay Brandner,\textsuperscript{1} Keiji Saito,\textsuperscript{2} and Udo Seifert\textsuperscript{1}}

\affiliation{\textsuperscript{{{\rm 1}}}II. Institut f\"ur Theoretischen Physik, Universit\"at Stuttgart, 70550 Stuttgart, Germany\\
\textsuperscript{{{\rm 2}}}Department of Physics, Keio University, 3-14-1 Hiyoshi, Kohoku-ku, Yokohama, Japan 223-8522}

%\thanks{}


\date{\today}

\begin{abstract}

For thermoelectric transport in the presence of a magnetic field 
that breaks time-reversal symmetry, a strong bound on the Onsager coefficients 
is derived within a general set-up using three terminals. Asymmetric 
Onsager coefficients lead to a maximum efficiency
substantially smaller than the Carnot efficiency reaching only
$\eta_C/4$ in the limit of strong asymmetry. Related bounds are derived
for efficiency at maximum power, which can become larger than the 
Curzon-Ahlborn value $\eta_C/2$, and for a cooling device.  Our approach 
reveals that in the presence of reversible currents the standard analysis 
based on the positivity of entropy production is incomplete without 
considering the role of current conservation explicitly.\\

PACS: numbers:  05.70.Ln, 72.15.Jf

\end{abstract}

\pacs{}
%\keywords{}

\maketitle



\def\dbar{{\mathchar'26\mkern-12mu d}} 

The Onsager reciprocity theorem derived more than
eighty years ago is arguably the most important constraint for the phenomenological 
framework of linear irreversible thermodynamics \cite{Onsager1931}. In its original 
form it states that the linear kinetic coefficient $L_{ik}(\mathbf{B})$, which relates the 
affinity $\mathcal{F}_k$ to the flux $J_i$ is equal to the reciprocal coefficient 
$L_{ki}(-\mathbf{B})$ provided both $i$ and $k$ refer to quantities with even signature 
under time-reversal  \cite{Callen1985}. The constant magnetic field $\mathbf{B}$ breaks
the time-reversal symmetry.
A generalization which includes variables with odd signature was given by Casimir 
in 1948 \cite{Casimir1945}. Ever since, the reciprocal relations have turned out to 
be extremely useful for a plethora of applications. A prominent one is the unified 
theory of thermoelectricity which allows to treat the various thermoelectric effects
such as the Peltier, the Seebeck or the Thomson effect on an equal footing thus 
revealing their interdependencies \cite{Goupil2011,Callen1985}. 

It is well known that properly designed energy filters allow a substantial enhancement of 
the efficiency of thermoelectric devices \cite{Mahan1996}. Linke and 
co-workers \cite{Humphrey2002, Humphrey2005}
showed that, in principle, even Carnot efficiency is attainable.
Despite these promising theoretical results, the 
actual performance of present devices is notoriously much more modest, 
which keeps the search for better thermoelectric materials a very active and increasingly
important research field
as reviewed in \cite{Vineis2010,Bell2008,Snyder2008,Dresselhaus2007}. Recently, in an
intriguing paper,
Benenti et al. pointed out that the presence of a magnetic field 
could in principle enhance the performance of a thermoelectric device \cite{Benenti2011}. 
In fact, their quite general thermodynamic analysis invoking only the Onsager 
reciprocity relations and the positivity of entropy production suggested the possibility 
of a device delivering finite power while still reaching Carnot efficiency.
Such a spectacular option deserves both further scrutiny and a search for a 
microscopic realization. In this Letter, we show that unitarity of the scattering 
matrix as a general physical principle imposes a strong restriction on
the Onsager coefficients that lead to a significant reduction of the attainable
efficiency of any thermoelectric device within the broad and well-established class 
of three-terminal models \cite{Horvat2012, Entin-Wohlman2012, Sanchez2011, Saito2011}.

\begin{figure}[b]
\epsfig{file=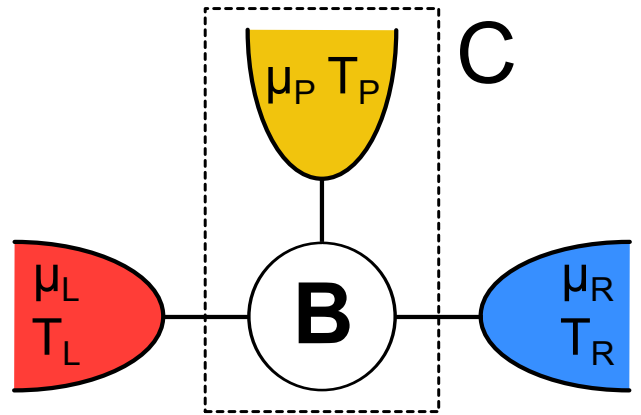,scale=2.0}
\caption{Sketch of a thermoelectric device in the presence of a magnetic field $\mathbf{B}$.
The entire dashed box represents the conductor C, which is connected to two reservoirs.
For the special case of the three terminal model, the conductor consists essentially of a scattering 
region and an additional probe terminal. \label{Heat Engine}}
\end{figure}

Thermoelectric transport can conveniently be discussed within the model sketched in
Fig.\ref{Heat Engine}. Two particle reservoirs of respective temperatures
$T_{{{\rm L}}} > T_{{{\rm R}}}$ and chemical potentials $\mu_{{{\rm L}}}< \mu_{{{\rm R}}}$ 
are connected by a conductor C, which allows for the exchange of heat and particles. 
Consequently, as soon as the steady state is reached, constant heat and particle currents,
$J_{q}$ and $J_{\rho}$, flow between the left and the right reservoirs. 
Depending on the sign of these currents, the machine works either as a power generator 
or a refrigerator. We now assume that both the temperature difference 
$\Delta T \equiv T_{{{\rm L}}} -T_{{{\rm R}}}>0$ and the chemical potential difference
$\Delta \mu \equiv \mu_{{{\rm L}}} - \mu_{{{\rm R}}} < 0$ are small compared 
to the respective reference values, which we choose to be $T\equiv T_{{{\rm R}}}$ and 
$\mu \equiv \mu_{{{\rm R}}} $. In this linear response regime,
the currents $J_{\rho}$ and $J_q$ are related to the affinities
$\mathcal{F}_{\rho} \equiv \Delta \mu / T$ and $\mathcal{F}_q \equiv \Delta T / T^2$
via the phenomenological equations
\begin{equation}
\begin{split}
J_{\rho} &= L_{\rho \rho} \mathcal{F}_{\rho} + L_{\rho q} \mathcal{F}_q\\
J_q & = L_{q \rho} \mathcal{F}_{\rho} + L_{qq} \mathcal{F}_q. 
\end{split}
\end{equation}
The constant rate of entropy production accompanying this transport process reads
\begin{equation}\label{Entropy Production Rate}
\dot{S} \! = \! \mathcal{F}_{\rho} J_{\rho} + \mathcal{F}_qJ_q \! 
= \! L_{\rho \rho} \mathcal{F}_{\rho}^2 + L_{qq} \mathcal{F}_q^2 
+ (L_{\rho q} + L_{q \rho}) \mathcal{F}_{\rho} \mathcal{F}_q. 
\end{equation}
Clearly, the features of this machine are completely determined by the kinetic coefficients
 $L_{ij}$ $(i,j=\rho,q)$, which are subject to two fundamental constraints. 
First, the second law requires $\dot{S}\geq 0$, which is equivalent to the conditions
\begin{equation}\label{Second Law}
L_{\rho \rho}, L_{qq} \geq 0 \quad \text{and} \quad L_{\rho \rho}L_{qq} 
- (L_{\rho q} + L_{q \rho})^2 / 4 \geq 0. 
\end{equation}
Second, Onsager's theorem implies the symmetry relation 
\begin{equation}\label{Onsager Symmetry}
L_{\rho q}(\mathbf{B}) = L_{q \rho}(-\mathbf{B}).
\end{equation} 
Here, we have reintroduced the magnetic field $\mathbf{B}$ which was suppressed 
so far in order to keep the notation slim. 

In the light of the Onsager reciprocity relation (\ref{Onsager Symmetry}), 
the expression (\ref{Heat Engine}) for the total entropy production rate suggests
a natural splitting of each current $J_i$ into a reversible and an irreversible 
part defined by
\begin{equation}
J^{\text{rev}}_i \equiv \frac{L_{ij} - L_{ji}}{2} \mathcal{F}_j \! \quad \text{and} 
\! \quad J^{\text{irr}}_i \equiv L_{ii} \mathcal{F}_i + \frac{L_{ij} + L_{ji}}{2} \mathcal{F}_j.
\end{equation}
Obviously, $J^{\text{rev}}_i$ vanishes for $\mathbf{B} =0$, i.e. for unbroken 
time-reversal symmetry. However, once the latter is broken, although not contributing
to $\dot{S}$, the reversible currents can, in principle, become arbitrarily large, 
since besides (\ref{Second Law}) no further general relations between $L_{ij}(\mathbf{B})$
and $L_{ji}(\mathbf{B})$ are known. Such unconstraint reversible currents could ultimately 
give rise to the possibility of dissipationless transport noticed by Benenti et al. 
\cite{Benenti2011} as mentioned above. 

For progress investigating possible further constraints
the phenomenological set-up discussed so far must be made more specific. A paradigmatic
model for the conductor C is shown inside the dashed box in Fig. \ref{Heat Engine}.
It consists of a central scattering region, 
with a constant magnetic field $\mathbf{B}$, and a third electronic reservoir (terminal) P,
whose temperature and chemical potential are chosen such that no net exchange of particles
and heat with the rest of the system occurs. The scattering region is connected to all of
the reservoirs via perfect, one-dimensional, infinitely long leads. In order to keep the 
model as simple as possible, we assume moderately damped \cite{Humphrey2002}, non-interacting 
electrons, which means that no inelastic scattering events take place outside 
the reservoirs and the electrons are transferred coherently between them. This set-up constitutes
the quite general class of three-terminal models on which we base our subsequent analysis. 
We emphasize that the additional terminal P plays a crucial role since it is well known 
that in a purely coherent two terminal set-up the off-diagonal Onsager coefficients must be even
functions of the magnetic field \cite{Buttiker1988a} and hence there are no reversible currents.
Such probe terminals, whose temperature and chemical potential are chosen self-consistently, 
were originally proposed by B\"uttiker \cite{Buttiker1986} and have 
become a common tool to simulate inelastic events in an otherwise conservative system.

Since we now have to deal with three reservoirs, we need accordingly four affinities
$\mathcal{F}_{\rho}^A \equiv (\mu_A -\mu) / T$ and $\mathcal{F}_q^{A} \equiv ( T_A - T)/ T^2$ $(A=L,P)$,
where we still use the temperature and chemical potential of the right reservoir as reference
values $\mu$ and $T$. For convenience, we collect the affinities in two vectors
$\boldsymbol{\mathcal{F}}^A \equiv ( \mathcal{F}^A_{\rho}, \mathcal{F}_q^A)$. 
Analogously we define the current vectors $\mathbf{J}^A = ( e J_{\rho}^A, J_q^A)$,
with $J^A_{\rho}$ and $J^A_q$ the particle and heat currents flowing out of reservoir $A$
and $e$ the electronic unit charge.
This vector notation allows us to write the phenomenological equations in the rather compact form
\begin{equation}\label{4x4 System}
\binom{\mathbf{J}^L}{\mathbf{J}^P} = \mathbb{L}'
\binom{ \boldsymbol{\mathcal{F}}^L}{ \boldsymbol{\mathcal{F}}^P} \quad \text{with} \quad
\mathbb{L}' \equiv \left( \! \! \begin{array}{c c} \mathbb{L}_{LL}' & \mathbb{L}_{LP}'\\
\mathbb{L}_{PL}' & \mathbb{L}_{PP}' \end{array} \! \! \right).
\end{equation}
Here, the $\mathbb{L}_{AB}'$ $(A,B =L,P)$ are $2 \times 2$-matrices of kinetic coefficients. 
By virtue of the additional constraint $\mathbf{J}^P = 0$, we can moreover eliminate
$\boldsymbol{\mathcal{F}}^P$ and thereby reduce (\ref{4x4 System}) to a system of two equations
\begin{equation}
\mathbf{J}^L = \mathbb{L} \boldsymbol{\mathcal{F}}^L
\end{equation}
with an effective matrix of kinetic coefficients
\begin{equation}\label{Reduced Onsager Matrix}
\mathbb{L} \equiv \mathbb{L}_{LL}' - \mathbb{L}_{LP}' \mathbb{L}_{PP}'^{-1} \mathbb{L}'_{PL}
 \equiv \left( \! \! \begin{array}{c c} L_{\rho \rho} & L_{\rho q} \\ L_{q \rho} & L_{qq}
\end{array} \! \! \right)
\end{equation}
relevant for the net currents from L to R.

Explicit expressions for the $2 \times 2$ block matrices $\mathbb{L}'_{AB}$ of
kinetic coefficients showing up in (\ref{4x4 System}) arise from the 
multi-terminal Landauer formula \cite{Butcher1990,Sivan1986} 
\begin{multline}\label{Landauer Formula}
\mathbb{L}_{AB}' = \frac{e^2 T }{ h} \int^{\infty}_{-\infty} dE \; F\left(E \right)
\left( \! \! \begin{array}{c c} 1  & \frac{E- \mu}{e}\\  \frac{E- \mu}{e} & 
\left( \frac{E- \mu}{e} \right)^2 \end{array} \! \! \right) \\ 
\times \left(\delta_{AB} - \left| S_{AB}(E, \mathbf{B}) \right|^2 \right),
\end{multline}
where
\begin{equation}
F(E) \equiv \left[ 4 k_B T \cosh^2 \left( \frac{E-\mu}{k_B T} \right) \right]^{-1}
\end{equation}
is the negative derivative of the Fermi function, $k_B$ denotes Boltzmann's constant 
and $h$ Planck's constant.
$S_{AB}(E, \mathbf{B}) \neq S_{BA}(E, \mathbf{B})$ are the
matrix elements of the $3 \times 3$ scattering matrix  $\mathbb{S}(E,\mathbf{B})$  that
describes the passage of electrons with energy $E$ through the scattering region. Current 
conservation requires that $\mathbb{S}(E,\mathbf{B})$ is unitary and the time reversal 
invariance of unitary dynamics implies the symmetry 
$\mathbb{S}(E,\mathbf{B}) = \mathbb{S}(E, - \mathbf{B})^t$ \cite{Butcher1990}. 
We emphasize that in the presence of a magnetic field, this symmetry permits 
$\mathbb{S}(E, \mathbf{B})$ to be non symmetric.
Consequently, despite the fact that the $\mathbb{L}'_{AB}$ are still symmetric,
the reduced matrix (\ref{Reduced Onsager Matrix}) will generically acquire a non-vanishing 
asymmetric part as desired. 

We now derive a constraint on the asymmetry of the reduced matrix $\mathbb{L}$.
First, we define the Hermitian matrices 
\begin{equation}\label{K Matrix}
\mathbb{K}' \equiv (\mathbb{L}' + \mathbb{L}'^t) + i \sqrt{3}(\mathbb{L}' - \mathbb{L}'^t)
\end{equation}
and 
\begin{equation}\label{Reduced K Matrix}
\mathbb{K} \equiv (\mathbb{L} + \mathbb{L}^t) + i \sqrt{3}(\mathbb{L} - \mathbb{L}^t) 
\equiv \left(\! \! \begin{array}{c c} K_{\rho \rho} & K_{\rho q} \\ K_{ \rho q}^{\ast} & K_{qq}
\end{array} \! \! \right),
\end{equation}
where $\mathbb{L}'$ is the full $4 \times 4$ Onsager-matrix introduced in (\ref{4x4 System}),
while $\mathbb{L}$ is the reduced matrix (\ref{Reduced Onsager Matrix}). 
Second, as a consequence of the unitarity of the scattering matrix
$\mathbb{S}(E,\textbf{B})$, $\mathbb{K}'$ can be shown to be positive semidefinite on $\mathbb{C}^4$ \cite{SI}.  
Third, since for any $\mathbf{z} \in \mathbb{C}^2$ 
\begin{equation}
\mathbf{z}^{\dagger} \mathbb{K} \mathbf{z} = \mathbf{z}'^{\dagger} \mathbb{K}'\mathbf{z}' \geq 0  
\; \; \; \text{with}  \; \; \;  \mathbf{z}' 
\equiv \binom{\mathbf{z}}{- {\mathbb{L}'}_{PP}^{-1}  \mathbb{L}_{PL}' \mathbf{z} },
\end{equation}
it then follows that the reduced matrix $\mathbb{K}$ must be positive semidefinite
on $\mathbb{C}^2$. 
Therefore the matrix elements of $\mathbb{K}$ have to obey the inequalities
\begin{equation}\label{Bound on Matrix Elements of Reduced K Matrix}
K_{\rho \rho}, K_{qq} \geq 0 \quad \text{and} 
\quad K_{\rho \rho} K_{qq} - K_{\rho q} K_{\rho q}^{\ast} \geq 0.
\end{equation}
We note that the $K_{ij}$ $(i,j=\rho,q)$ are rather complicated functions of the matrix elements
of the full Onsager-matrix $\mathbb{L}'$, which would make it a quite challenging task to obtain 
(\ref{Bound on Matrix Elements of Reduced K Matrix}) directly from (\ref{Reduced Onsager Matrix})
and (\ref{Landauer Formula}). If we insert the definition (\ref{Reduced K Matrix})  
of the $K_{ij}$ in terms of the $L_{ij}$, we immediately get the relation
\begin{equation}\label{New Bound}
L_{\rho \rho} L_{qq} + L_{\rho q} L_{q \rho} - L_{\rho q}^2 - L_{q \rho}^2 \geq 0,
\end{equation}
which constitutes our first main result. We emphasize that (\ref{New Bound}) provides a much 
stronger constraint than (\ref{Second Law}) since the former can be rewritten as
\begin{equation}
L_{\rho \rho} L_{qq} - (L_{\rho q} + L_{q \rho})^2 / 4 \geq 3(L_{\rho q} - L_{q \rho})^2 / 4.
\end{equation} 
Indeed, we recover (\ref{Second Law}) only if $L_{\rho q}$ is equal to $L_{q \rho}$. 
As soon as the kinetic coefficients contain finite asymmetric parts, the left hand side of 
(\ref{New Bound}) must be strictly larger than zero. In other words, the reversible currents
associated with the asymmetric part of $\mathbb{L}$ come at the price of a stronger lower
bound on the Onsager coefficients and hence on the entropy production rate 
(\ref{Entropy Production Rate}) than the bare second law (\ref{Second Law}) requires.

\begin{figure}
\epsfig{file=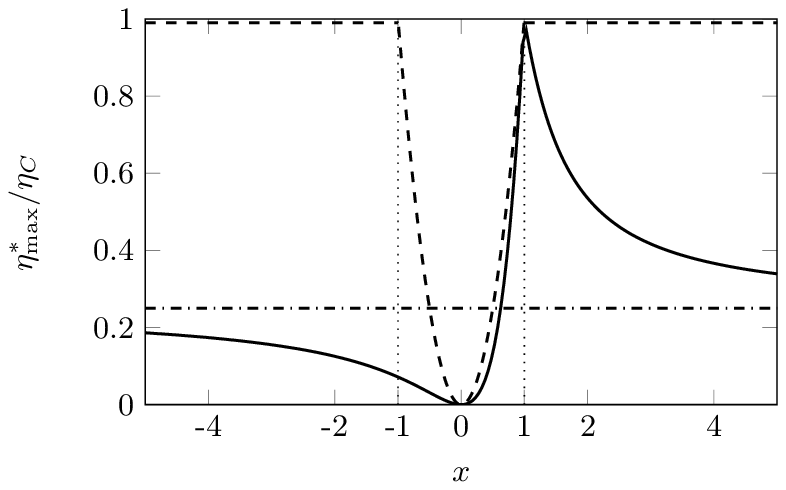}
\epsfig{file=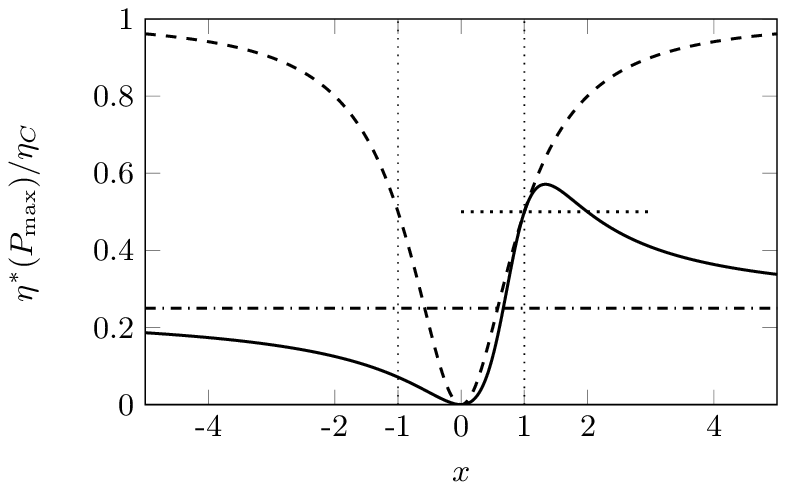}
\caption{Bounds on benchmarks in units of $\eta_C$ as functions of the asymmetry parameter $x$.
The upper diagram shows $\eta_{\text{max}}^{\ast}$, the lower one $\eta(P_{\text{max}})^{\ast}$. 
The solid curves represent the bounds following from relation (\ref{New Bound}). In the limit 
$x \rightarrow \pm \infty$ both functions asymptotically approach the value $1/4$, 
which is shown by the dash-dotted lines. For comparison, the bounds obtained by Benenti et al. 
\cite{Benenti2011} solely from the second law (\ref{Second Law}) have been included as dashed curves. 
The dotted line in the lower panel indicates the Curzon-Ahlborn limit $1/2$.
\label{Diagrams Power Generator}}
\end{figure}

The new bound (\ref{New Bound}) has profound consequences  for the performance of the model
as a thermoelectric heat engine. Following the lines of Benenti et al. \cite{Benenti2011}, we
introduce the dimensionless parameters
\begin{equation}\label{Dimensionless Parameters}
y \equiv L_{\rho q} L_{q \rho} / \text{Det} \; \mathbb{L} \quad \text{and} 
\quad x \equiv L_{\rho q} / L_{q \rho}.
\end{equation}
Expressed in terms of $x$ and $y$, the inequality (\ref{New Bound}) reads
\begin{equation}\label{Bound in Dimensionless Parameters}
\begin{split}
h(x) \leq 4y \leq 0  \quad \text{if} \quad x<0,\\
0 \leq 4y \leq h(x)  \quad \text{if} \quad x>0,
\end{split}
\end{equation}
where $h(x) \equiv 4x/(x-1)^2$ .
The most important benchmarks for the performance of a heat engine - namely the maximum efficiency
$\eta_{\text{max}}$ and the efficiency at maximum power $\eta(P_{\text{max}})$ - admit the 
analytic expressions \cite{Benenti2011}
\begin{equation}
\eta_{\text{max}} = \eta_C x \frac{\sqrt{y+1}-1}{\sqrt{y+1}+1} \;  \text{and}  \; 
\eta(P_{\text{max}}) = \eta_C \frac{xy}{4+ 2y}.
\end{equation} 

Here, we have denoted by $ \eta_C = 1- T_{{{\rm R}}}/T_{{{\rm L}}} \approx T \mathcal{F}^{L}_q $ 
the Carnot efficiency, which is the absolute upper bound for the attainable efficiency
following from the second law.  Both benchmarks become maximal
for $4y=h(x)$. The resulting bounds, $\eta^{\ast}_{\text{max}}$ and $\eta^{\ast}(P_{\text{max}})$,
which are our second main result, are plotted in Fig. \ref{Diagrams Power Generator}. It shows
how the maximum efficiency decays rapidly as the asymmetry parameter $x$ deviates from its symmetric
value $1$. In the limit  $x \rightarrow \pm \infty$, maximum efficiency and efficiency
at maximum power both approach $\eta_C / 4$. This result implies essentially that from the perspective
of maximally attainable efficiency the thermodynamic cost of the reversible currents is larger
than the benefit they bring. For efficiency at maximum power, the situation is somewhat different.
If $x$ is only slightly larger than $1$, the Curzon-Ahlborn-limit $\eta_{CA} = \eta_C /2$ 
\cite{Curzon1975,VandenBroeck2005,Esposito2010,Seifert2011,VandenBroeck2012}, reached for $x=1$, can be overcome in a small range of $x$ values with 
a maximum of $4 \eta_C/7$ at $x=4/3$. 
This result shows that despite the strong bounds on $\eta_{\text{max}}$ it may in principle 
be possible to improve the performance of the machine by breaking the time-reversal symmetry
slightly.

\begin{figure}
\epsfig{file=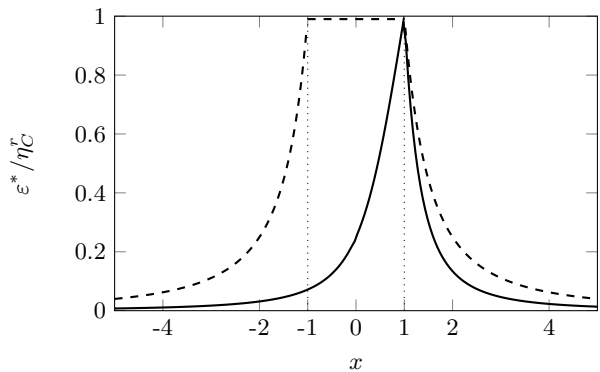}
\caption{Bounds on the coefficient of performance of a refrigerator 
$\varepsilon$ in units of $\eta_C^r$ as a function
of the asymmetry parameter $x$. The dashed line follows from bare second law, the solid line from 
the stronger relation (\ref{New Bound}).\label{Diagram Refrigerator}}
\end{figure}

Finally, we discuss the consequence of the bound (\ref{New Bound}) for  the model as a refrigerator.
In this case the most im\-portant benchmark is the coefficient of performance 
$\varepsilon \equiv - J^{L}_q / T \mathcal{F}^L_{\rho} J^{L}_{\rho}$ \cite{Callen1985} defined 
as the ratio of the heat current extracted from the cold reservoir and the absorbed power.
In terms of the dimensionless parameters (\ref{Dimensionless Parameters}) $\varepsilon$ reads \cite{Benenti2011}
\begin{equation}
\varepsilon = \frac{\eta_C^r}{x} \frac{\sqrt{y+1}-1}{\sqrt{y+1}+1},
\end{equation}
where $\eta_C^r = T_{{{\rm R}}}/ (T_{{{\rm L}}} -T_{{{\rm R}}}) \approx 1/T \mathcal{F}^L_q$ is the 
efficiency of an ideal refrigerator. 
Like for $\eta_{\text{max}}$ and $\eta(P_{\text{max}})$, the maximum 
$\varepsilon^{\ast}$ of $\varepsilon$ is attained for $4y=h(x)$. The resulting bound is plotted
in Fig. \ref{Diagram Refrigerator}. Again we observe that the efficiency deteriorates rapidly as 
$x$ deviates from $1$.

In conclusion, we have investigated the thermoelectric transport
properties of the most general version of the paradigmatic three terminal model in the presence of broken
time reversal symmetry. We have derived a strong constraint on the linear transport coefficients
that can be obtained neither from Onsager's principle of microreversibility nor from thermodynamic arguments
invoking only the second law. This constraint implies strong bounds on both maximum efficiency
and efficiency at maximum power if this device is operated as a thermoelectric heat engine or refrigerator. 

We emphasize that our results rely solely on the unitarity of the scattering matrix.
Therefore our new relation (\ref{New Bound}) applies to any specific model which can be formulated within
the three terminal set-up,  including especially the ones based on quantum dots \cite{Saito2011} and ballistic
microjunctions \cite{Sanchez2011}, which have been proposed in this context recently. It might even apply to a model
introduced by Entin-Wohlman and Aharony \cite{Entin-Wohlman2012}, which includes a phonon bath locally interacting
with electrons transferred between the reservoirs. Furthermore it is noteworthy
that by following exactly the same lines the constraint (\ref{New Bound}) can be shown to apply to the three 
terminal railway switch transport model \cite{Horvat2012}. 
This model may be regarded as a classical analogue of the quantum three terminal model discussed here. 
Indeed, our considerations are not restricted to the quantum realm, since ultimately the unitarity of the 
scattering matrix is nothing but the manifestation of the law of current conservation, which should be considered
the fundamental principle underlying our bound. 
 
Despite this generality, the question whether a similar bound can be found for more complex model classes,
including for example a larger number of probe terminals or models with genuinely interacting electrons, 
remains open and should constitute an important subject for further investigations.\\

\begin{acknowledgments}
Acknowledgments: K.B. and U.S. acknowledge support from ESF through
the EPSD network. K.S. acknowledges support from MEXT (23740289).
\end{acknowledgments}

% Create the reference section using BibTeX




\begin{thebibliography}{26}%
\makeatletter
\providecommand \@ifxundefined [1]{%
 \@ifx{#1\undefined}
}%
\providecommand \@ifnum [1]{%
 \ifnum #1\expandafter \@firstoftwo
 \else \expandafter \@secondoftwo
 \fi
}%
\providecommand \@ifx [1]{%
 \ifx #1\expandafter \@firstoftwo
 \else \expandafter \@secondoftwo
 \fi
}%
\providecommand \natexlab [1]{#1}%
\providecommand \enquote  [1]{``#1''}%
\providecommand \bibnamefont  [1]{#1}%
\providecommand \bibfnamefont [1]{#1}%
\providecommand \citenamefont [1]{#1}%
\providecommand \href@noop [0]{\@secondoftwo}%
\providecommand \href [0]{\begingroup \@sanitize@url \@href}%
\providecommand \@href[1]{\@@startlink{#1}\@@href}%
\providecommand \@@href[1]{\endgroup#1\@@endlink}%
\providecommand \@sanitize@url [0]{\catcode `\\12\catcode `\$12\catcode
  `\&12\catcode `\#12\catcode `\^12\catcode `\_12\catcode `\%12\relax}%
\providecommand \@@startlink[1]{}%
\providecommand \@@endlink[0]{}%
\providecommand \url  [0]{\begingroup\@sanitize@url \@url }%
\providecommand \@url [1]{\endgroup\@href {#1}{\urlprefix }}%
\providecommand \urlprefix  [0]{URL }%
\providecommand \Eprint [0]{\href }%
\providecommand \doibase [0]{http://dx.doi.org/}%
\providecommand \selectlanguage [0]{\@gobble}%
\providecommand \bibinfo  [0]{\@secondoftwo}%
\providecommand \bibfield  [0]{\@secondoftwo}%
\providecommand \translation [1]{[#1]}%
\providecommand \BibitemOpen [0]{}%
\providecommand \bibitemStop [0]{}%
\providecommand \bibitemNoStop [0]{.\EOS\space}%
\providecommand \EOS [0]{\spacefactor3000\relax}%
\providecommand \BibitemShut  [1]{\csname bibitem#1\endcsname}%
\let\auto@bib@innerbib\@empty
%</preamble>
\bibitem [{\citenamefont {Onsager}(1931)}]{Onsager1931}%
  \BibitemOpen
  \bibfield  {author} {\bibinfo {author} {\bibfnamefont {L.}~\bibnamefont
  {Onsager}},\ }\href@noop {} {\bibfield  {journal} {\bibinfo  {journal} {Phys.
  Rev.}\ }\textbf {\bibinfo {volume} {38}},\ \bibinfo {pages} {2265} (\bibinfo
  {year} {1931})}\BibitemShut {NoStop}%
\bibitem [{\citenamefont {Callen}(1985)}]{Callen1985}%
  \BibitemOpen
  \bibfield  {author} {\bibinfo {author} {\bibfnamefont {H.~B.}\ \bibnamefont
  {Callen}},\ }\href@noop {} {\emph {\bibinfo {title} {{Thermodynamics and an
  Introduction to Thermostatics}}}},\ \bibinfo {edition} {2nd}\ ed.\ (\bibinfo
  {publisher} {John Wiley \& Sons},\ \bibinfo {address} {New York},\ \bibinfo
  {year} {1985})\BibitemShut {NoStop}%
\bibitem [{\citenamefont {Casimir}(1945)}]{Casimir1945}%
  \BibitemOpen
  \bibfield  {author} {\bibinfo {author} {\bibfnamefont {H.~B.~G.}\
  \bibnamefont {Casimir}},\ }\href@noop {} {\bibfield  {journal} {\bibinfo
  {journal} {Rev. Mod. Phys.}\ }\textbf {\bibinfo {volume} {17}},\ \bibinfo
  {pages} {343} (\bibinfo {year} {1945})}\BibitemShut {NoStop}%
\bibitem [{\citenamefont {Goupil}\ \emph {et~al.}(2011)\citenamefont {Goupil},
  \citenamefont {Seifert}, \citenamefont {Zabrocki}, \citenamefont
  {M\"{u}ller},\ and\ \citenamefont {Snyder}}]{Goupil2011}%
  \BibitemOpen
  \bibfield  {author} {\bibinfo {author} {\bibfnamefont {C.}~\bibnamefont
  {Goupil}}, \bibinfo {author} {\bibfnamefont {W.}~\bibnamefont {Seifert}},
  \bibinfo {author} {\bibfnamefont {K.}~\bibnamefont {Zabrocki}}, \bibinfo
  {author} {\bibfnamefont {E.}~\bibnamefont {M\"{u}ller}}, \ and\ \bibinfo
  {author} {\bibfnamefont {G.~J.}\ \bibnamefont {Snyder}},\ }\href {\doibase
  10.3390/e13081481} {\bibfield  {journal} {\bibinfo  {journal} {Entropy}\
  }\textbf {\bibinfo {volume} {13}},\ \bibinfo {pages} {1481} (\bibinfo {year}
  {2011})}\BibitemShut {NoStop}%
\bibitem [{\citenamefont {Mahan}\ and\ \citenamefont {Sofo}(1996)}]{Mahan1996}%
  \BibitemOpen
  \bibfield  {author} {\bibinfo {author} {\bibfnamefont {G.~D.}\ \bibnamefont
  {Mahan}}\ and\ \bibinfo {author} {\bibfnamefont {J.~O.}\ \bibnamefont
  {Sofo}},\ }\href
  {http://www.pubmedcentral.nih.gov/articlerender.fcgi?artid=38761\&tool=pmcentrez\&rendertype=abstract}
  {\bibfield  {journal} {\bibinfo  {journal} {Proc. Natl. Acad. Sci. USA}\
  }\textbf {\bibinfo {volume} {93}},\ \bibinfo {pages} {7436} (\bibinfo {year}
  {1996})}\BibitemShut {NoStop}%
\bibitem [{\citenamefont {Humphrey}\ \emph {et~al.}(2002)\citenamefont
  {Humphrey}, \citenamefont {Newbury}, \citenamefont {Taylor},\ and\
  \citenamefont {Linke}}]{Humphrey2002}%
  \BibitemOpen
  \bibfield  {author} {\bibinfo {author} {\bibfnamefont {T.~E.}~\bibnamefont
  {Humphrey}}, \bibinfo {author} {\bibfnamefont {R.}~\bibnamefont {Newbury}},
  \bibinfo {author} {\bibfnamefont {R.~P.}~\bibnamefont {Taylor}}, \ and\ \bibinfo
  {author} {\bibfnamefont {H.}~\bibnamefont {Linke}},\ }\href {\doibase
  10.1103/PhysRevLett.89.116801} {\bibfield  {journal} {\bibinfo  {journal}
  {Phys. Rev. Lett.}\ }\textbf {\bibinfo {volume} {89}},\ \bibinfo {pages}
  {116801} (\bibinfo {year} {2002})}\BibitemShut {NoStop}%
\bibitem [{\citenamefont {Humphrey}\ and\ \citenamefont
  {Linke}(2005)}]{Humphrey2005}%
  \BibitemOpen
  \bibfield  {author} {\bibinfo {author} {\bibfnamefont {T.~E.}\ \bibnamefont
  {Humphrey}}\ and\ \bibinfo {author} {\bibfnamefont {H.}~\bibnamefont
  {Linke}},\ }\href {\doibase 10.1103/PhysRevLett.94.096601} {\bibfield
  {journal} {\bibinfo  {journal} {Phys. Rev. Lett.}\ }\textbf {\bibinfo
  {volume} {94}},\ \bibinfo {pages} {096601} (\bibinfo {year}
  {2005})}\BibitemShut {NoStop}%
\bibitem [{\citenamefont {Vineis}\ \emph {et~al.}(2010)\citenamefont {Vineis},
  \citenamefont {Shakouri}, \citenamefont {Majumdar},\ and\ \citenamefont
  {Kanatzidis}}]{Vineis2010}%
  \BibitemOpen
  \bibfield  {author} {\bibinfo {author} {\bibfnamefont {C.~J.}\ \bibnamefont
  {Vineis}}, \bibinfo {author} {\bibfnamefont {A.}~\bibnamefont {Shakouri}},
  \bibinfo {author} {\bibfnamefont {A.}~\bibnamefont {Majumdar}}, \ and\
  \bibinfo {author} {\bibfnamefont {M.~G.}\ \bibnamefont {Kanatzidis}},\ }\href
  {\doibase 10.1002/adma.201000839} {\bibfield  {journal} {\bibinfo  {journal}
  {Adv. Mat.}\ }\textbf {\bibinfo {volume} {22}},\ \bibinfo {pages} {3970}
  (\bibinfo {year} {2010})}\BibitemShut {NoStop}%
\bibitem [{\citenamefont {Bell}(2008)}]{Bell2008}%
  \BibitemOpen
  \bibfield  {author} {\bibinfo {author} {\bibfnamefont {L.~E.}\ \bibnamefont
  {Bell}},\ }\href {\doibase 10.1126/science.1158899} {\bibfield  {journal}
  {\bibinfo  {journal} {Science}\ }\textbf {\bibinfo {volume} {321}},\ \bibinfo
  {pages} {1457} (\bibinfo {year} {2008})}\BibitemShut {NoStop}%
\bibitem [{\citenamefont {Snyder}\ and\ \citenamefont
  {Toberer}(2008)}]{Snyder2008}%
  \BibitemOpen
  \bibfield  {author} {\bibinfo {author} {\bibfnamefont {G.~J.}\ \bibnamefont
  {Snyder}}\ and\ \bibinfo {author} {\bibfnamefont {S.}~\bibnamefont
  {Toberer}},\ }\href@noop {} {\bibfield  {journal} {\bibinfo  {journal}
  {Nature Mater.}\ }\textbf {\bibinfo {volume} {7}},\ \bibinfo {pages} {105}
  (\bibinfo {year} {2008})}\BibitemShut {NoStop}%
\bibitem [{\citenamefont {Dresselhaus}\ \emph {et~al.}(2007)\citenamefont
  {Dresselhaus}, \citenamefont {Chen}, \citenamefont {Tang}, \citenamefont
  {Yang}, \citenamefont {Lee},\ and\ \citenamefont {Wang}}]{Dresselhaus2007}%
  \BibitemOpen
  \bibfield  {author} {\bibinfo {author} {\bibfnamefont {B.~M.~S.}\
  \bibnamefont {Dresselhaus}}, \bibinfo {author} {\bibfnamefont
  {G.}~\bibnamefont {Chen}}, \bibinfo {author} {\bibfnamefont {M.~Y.}\
  \bibnamefont {Tang}}, \bibinfo {author} {\bibfnamefont {R.}~\bibnamefont
  {Yang}}, \bibinfo {author} {\bibfnamefont {H.}~\bibnamefont {Lee}}, \ and\
  \bibinfo {author} {\bibfnamefont {D.}~\bibnamefont {Wang}},\ }\href {\doibase
  10.1002/adma.200600527} {\bibfield  {journal} {\bibinfo  {journal} {Adv.
  Mat.}\ }\textbf {\bibinfo {volume} {19}},\ \bibinfo {pages} {1043} (\bibinfo
  {year} {2007})}\BibitemShut {NoStop}%
\bibitem [{\citenamefont {Benenti}\ \emph {et~al.}(2011)\citenamefont
  {Benenti}, \citenamefont {Saito},\ and\ \citenamefont
  {Casati}}]{Benenti2011}%
  \BibitemOpen
  \bibfield  {author} {\bibinfo {author} {\bibfnamefont {G.}~\bibnamefont
  {Benenti}}, \bibinfo {author} {\bibfnamefont {K.}~\bibnamefont {Saito}}, \
  and\ \bibinfo {author} {\bibfnamefont {G.}~\bibnamefont {Casati}},\ }\href
  {\doibase 10.1103/PhysRevLett.106.230602} {\bibfield  {journal} {\bibinfo
  {journal} {Phys. Rev. Lett.}\ }\textbf {\bibinfo {volume} {106}},\ \bibinfo
  {pages} {230602} (\bibinfo {year} {2011})}\BibitemShut {NoStop}%
\bibitem [{\citenamefont {Horvat}\ \emph {et~al.}(2012)\citenamefont {Horvat},
  \citenamefont {Benenti},\ and\ \citenamefont {Casati}}]{Horvat2012}%
  \BibitemOpen
  \bibfield  {author} {\bibinfo {author} {\bibfnamefont {M.}~\bibnamefont
  {Horvat}}, \bibinfo {author} {\bibfnamefont {G.}~\bibnamefont {Benenti}}, \
  and\ \bibinfo {author} {\bibfnamefont {G.}~\bibnamefont {Casati}},\
  }\href@noop {} {\  (\bibinfo {year} {2012})},\ \Eprint
  {http://arxiv.org/abs/1207.6014v1} {arXiv:1207.6014v1} \BibitemShut {NoStop}%
\bibitem [{\citenamefont {Entin-Wohlman}\ and\ \citenamefont
  {Aharony}(2012)}]{Entin-Wohlman2012}%
  \BibitemOpen
  \bibfield  {author} {\bibinfo {author} {\bibfnamefont {O.}~\bibnamefont
  {Entin-Wohlman}}\ and\ \bibinfo {author} {\bibfnamefont {A.}~\bibnamefont
  {Aharony}},\ }\href {\doibase 10.1103/PhysRevB.85.085401} {\bibfield
  {journal} {\bibinfo  {journal} {Phys. Rev. B}\ }\textbf {\bibinfo {volume}
  {85}},\ \bibinfo {pages} {085401} (\bibinfo {year} {2012})}\BibitemShut
  {NoStop}%
\bibitem [{\citenamefont {S\'{a}nchez}\ and\ \citenamefont
  {Serra}(2011)}]{Sanchez2011}%
  \BibitemOpen
  \bibfield  {author} {\bibinfo {author} {\bibfnamefont {D.}~\bibnamefont
  {S\'{a}nchez}}\ and\ \bibinfo {author} {\bibfnamefont {L.}~\bibnamefont
  {Serra}},\ }\href {\doibase 10.1103/PhysRevB.84.201307} {\bibfield  {journal}
  {\bibinfo  {journal} {Phys. Rev. B}\ }\textbf {\bibinfo {volume} {84}},\
  \bibinfo {pages} {201307(R)} (\bibinfo {year} {2011})}\BibitemShut {NoStop}%
\bibitem [{\citenamefont {Saito}\ \emph {et~al.}(2011)\citenamefont {Saito},
  \citenamefont {Benenti}, \citenamefont {Casati},\ and\ \citenamefont
  {Prosen}}]{Saito2011}%
  \BibitemOpen
  \bibfield  {author} {\bibinfo {author} {\bibfnamefont {K.}~\bibnamefont
  {Saito}}, \bibinfo {author} {\bibfnamefont {G.}~\bibnamefont {Benenti}},
  \bibinfo {author} {\bibfnamefont {G.}~\bibnamefont {Casati}}, \ and\ \bibinfo
  {author} {\bibfnamefont {T.}~\bibnamefont {Prosen}},\ }\href {\doibase
  10.1103/PhysRevB.84.201306} {\bibfield  {journal} {\bibinfo  {journal} {Phys.
  Rev. B}\ }\textbf {\bibinfo {volume} {84}},\ \bibinfo {pages} {201306(R)}
  (\bibinfo {year} {2011})}\BibitemShut {NoStop}%
\bibitem [{\citenamefont {B\"{u}ttiker}(1988)}]{Buttiker1988a}%
  \BibitemOpen
  \bibfield  {author} {\bibinfo {author} {\bibfnamefont {M.}~\bibnamefont
  {B\"{u}ttiker}},\ }\href@noop {} {\bibfield  {journal} {\bibinfo  {journal}
  {IBM J. Res. Develop.}\ }\textbf {\bibinfo {volume} {32}},\ \bibinfo {pages}
  {317} (\bibinfo {year} {1988})}\BibitemShut {NoStop}%
\bibitem [{\citenamefont {B\"{u}ttiker}(1986)}]{Buttiker1986}%
  \BibitemOpen
  \bibfield  {author} {\bibinfo {author} {\bibfnamefont {M.}~\bibnamefont
  {B\"{u}ttiker}},\ }\href@noop {} {\bibfield  {journal} {\bibinfo  {journal}
  {Phys. Rev. B}\ }\textbf {\bibinfo {volume} {33}},\ \bibinfo {pages} {3020}
  (\bibinfo {year} {1986})}\BibitemShut {NoStop}%
\bibitem [{\citenamefont {Butcher}(1990)}]{Butcher1990}%
  \BibitemOpen
  \bibfield  {author} {\bibinfo {author} {\bibfnamefont {P.~N.}\ \bibnamefont
  {Butcher}},\ }\href@noop {} {\bibfield  {journal} {\bibinfo  {journal} {J.
  Phys. Condens. Matter}\ }\textbf {\bibinfo {volume} {2}},\ \bibinfo {pages}
  {4869} (\bibinfo {year} {1990})}\BibitemShut {NoStop}%
\bibitem [{\citenamefont {Sivan}\ and\ \citenamefont {Imry}(1986)}]{Sivan1986}%
  \BibitemOpen
  \bibfield  {author} {\bibinfo {author} {\bibfnamefont {U.}~\bibnamefont
  {Sivan}}\ and\ \bibinfo {author} {\bibfnamefont {Y.}~\bibnamefont {Imry}},\
  }\href@noop {} {\bibfield  {journal} {\bibinfo  {journal} {Phys. Rev. B}\
  }\textbf {\bibinfo {volume} {33}},\ \bibinfo {pages} {551} (\bibinfo {year}
  {1986})}\BibitemShut {NoStop}%
\bibitem [{SI()}]{SI}%
  \BibitemOpen
  \href@noop {} {}\bibinfo {note} {See Supplemental Material at [URL] for more
  details}\BibitemShut {NoStop}%
\bibitem [{\citenamefont {Curzon}\ and\ \citenamefont
  {Ahlborn}(1975)}]{Curzon1975}%
  \BibitemOpen
  \bibfield  {author} {\bibinfo {author} {\bibfnamefont {F.~L.}\ \bibnamefont
  {Curzon}}\ and\ \bibinfo {author} {\bibfnamefont {B.}~\bibnamefont
  {Ahlborn}},\ }\href {\doibase 10.1119/1.10023} {\bibfield  {journal}
  {\bibinfo  {journal} {Am. J. Phys.}\ }\textbf {\bibinfo {volume} {43}},\
  \bibinfo {pages} {22} (\bibinfo {year} {1975})}\BibitemShut {NoStop}%
\bibitem [{\citenamefont {{Van den Broeck}}(2005)}]{VandenBroeck2005}%
  \BibitemOpen
  \bibfield  {author} {\bibinfo {author} {\bibfnamefont {C.}~\bibnamefont {{Van
  den Broeck}}},\ }\href {\doibase 10.1103/PhysRevLett.95.190602} {\bibfield
  {journal} {\bibinfo  {journal} {Phys. Rev. Lett.}\ }\textbf {\bibinfo
  {volume} {95}},\ \bibinfo {pages} {190602} (\bibinfo {year}
  {2005})}\BibitemShut {NoStop}%
\bibitem [{\citenamefont {Esposito}\ \emph {et~al.}(2010)\citenamefont
  {Esposito}, \citenamefont {Kawai}, \citenamefont {Lindenberg},\ and\
  \citenamefont {{Van den Broeck}}}]{Esposito2010}%
  \BibitemOpen
  \bibfield  {author} {\bibinfo {author} {\bibfnamefont {M.}~\bibnamefont
  {Esposito}}, \bibinfo {author} {\bibfnamefont {R.}~\bibnamefont {Kawai}},
  \bibinfo {author} {\bibfnamefont {K.}~\bibnamefont {Lindenberg}}, \ and\
  \bibinfo {author} {\bibfnamefont {C.}~\bibnamefont {{Van den Broeck}}},\
  }\href {\doibase 10.1103/PhysRevLett.105.150603} {\bibfield  {journal}
  {\bibinfo  {journal} {Phys. Rev. Lett.}\ }\textbf {\bibinfo {volume} {105}},\
  \bibinfo {pages} {150603} (\bibinfo {year} {2010})}\BibitemShut {NoStop}%
\bibitem [{\citenamefont {Seifert}(2011)}]{Seifert2011}%
  \BibitemOpen
  \bibfield  {author} {\bibinfo {author} {\bibfnamefont {U.}~\bibnamefont
  {Seifert}},\ }\href {\doibase 10.1103/PhysRevLett.106.020601} {\bibfield
  {journal} {\bibinfo  {journal} {Phys. Rev. Lett.}\ }\textbf {\bibinfo
  {volume} {106}},\ \bibinfo {pages} {020601} (\bibinfo {year}
  {2011})}\BibitemShut {NoStop}%
\bibitem [{\citenamefont {{Van den Broeck}}\ \emph {et~al.}(2012)\citenamefont
  {{Van den Broeck}}, \citenamefont {Kumar},\ and\ \citenamefont
  {Lindenberg}}]{VandenBroeck2012}%
  \BibitemOpen
  \bibfield  {author} {\bibinfo {author} {\bibfnamefont {C.}~\bibnamefont {{Van
  den Broeck}}}, \bibinfo {author} {\bibfnamefont {N.}~\bibnamefont {Kumar}}, \
  and\ \bibinfo {author} {\bibfnamefont {K.}~\bibnamefont {Lindenberg}},\
  }\href {\doibase 10.1103/PhysRevLett.108.210602} {\bibfield  {journal}
  {\bibinfo  {journal} {Phys. Rev. Lett.}\ }\textbf {\bibinfo {volume} {108}},\
  \bibinfo {pages} {210602} (\bibinfo {year} {2012})}\BibitemShut {NoStop}%
\end{thebibliography}
\end{document}

\title{Supplemental Material for\\ Strong bounds on Onsager coefficients and efficiency\\for three terminal thermoelectric transport in a magnetic field}

\maketitle
\pagestyle{empty}

Here, we prove that the matrix
\begin{equation}
\mathbb{K}' \equiv (\mathbb{L}'+\mathbb{L}'^t) + i \sqrt{3}(\mathbb{L}'-\mathbb{L}'^t)
\end{equation}  
as introduced in Eq. (11) of the main text is positive semidefinite on $\mathbb{C}^4$. To this end, we first notice that by virtue of the multi-terminal Landauer formula [cf. Eq. (9) and Eq. (10) of the main text]
\begin{multline}\label{Landauer Formula}
\mathbb{L}_{AB}' =  \frac{e^2 T }{ h} \int_{-\infty}^{\infty} dE \; F\left(E \right)
\left( \! \! \begin{array}{c c} 1  & \frac{E- \mu}{e}\\  \frac{E- \mu}{e} & 
\left( \frac{E- \mu}{e} \right)^2 \end{array} \! \! \right) \\ 
\times \left(\delta_{AB} - \left| S_{AB}(E, \mathbf{B}) \right|^2 \right),
\end{multline}
the matrix elements $K'_{lm} \equiv (\mathbb{K}')_{lm}$ $(l,m= 1, \dots,4)$ can be written in the form
\begin{equation}
K'_{lm} =  \frac{e^2 T }{ h} \int_{-\infty}^{\infty}  dE \; F\left(E \right) \mathbf{v}_l^{\dagger}(E) \mathbb{V}(E, \mathbf{B}) \mathbf{v}_m(E)
\end{equation}
with
\begin{equation}\label{SI 1}
\mathbf{v}_1 \equiv \binom{1}{0}, \;  \mathbf{v}_2 \equiv \binom{\frac{E -\mu}{e}}{0}, \;  \mathbf{v}_3 \equiv \binom{0}{1}, \;  \mathbf{v}_4 \equiv \binom{0}{\frac{E -\mu}{e}}
\end{equation}
and the Hermitian matrix 
\begin{widetext}
\begin{equation}                                   
\mathbb{V} = \left( \! \! \begin{array}{c c}  2(1 - |S_{LL}|^2) \! &\!   |S_{LP}|^2 + |S_{PL}|^2 + i \sqrt{3}(|S_{LP}|^2 - |S_{PL}|^2)\\
                                        |S_{LP}|^2 + |S_{PL}|^2 - i \sqrt{3}(|S_{LP}|^2 - |S_{PL}|^2) \! & \! 2(1 - |S_{PP}|^2) \end{array} \! \! \right).                                     
\end{equation}
\end{widetext}
From (\ref{SI 1}) onwards we notationally suppress the explicit dependence of the $\mathbf{v}_l$, $\mathbb{V}$ and the elements of the scattering matrix
on energy $E$ and magnetic field $\mathbf{B}$.
For any $\mathbf{z} \equiv (z_1, z_2, z_3, z_4)^t \in \mathbb{C}^4$ we have
\begin{equation}\label{SI 2}
\begin{split}
\mathbf{z}^{\dagger} \mathbb{K}' \mathbf{z}  & = \sum_{l,m=1}^4 z_l^{\ast} z_m K'_{lm}\\
                                             & \hspace{-24pt} = \frac{e^2 T }{ h} \int_{-\infty}^{\infty} dE \; F(E) \left(  \sum_{l=1}^4 z_l^{\ast} \mathbf{v}^{\dagger}_l \right) \mathbb{V} 
                                                 \left(  \sum_{m=1}^4 z_m\mathbf{v}_m  \right)\\
                                             & \hspace{-24pt} = \frac{e^2 T }{ h} \int_{-\infty}^{\infty} dE \; F(E) \mathbf{y}^{\dagger} \mathbb{V} \mathbf{y}                                     \end{split}
\end{equation}
with
\begin{equation}
\mathbf{y} \equiv \sum_{m=1}^4 z_m \mathbf{v}_m.
\end{equation}
The last line of (\ref{SI 2}) reveals that if $\mathbb{V}$ is positive semidefinite, the same holds for $\mathbb{K}'$.

In order to show that $\mathbb{V}$ is indeed positive semidefinite, we first observe that both diagonal entries of
$\mathbb{V}$ are obviously non-negative. It therefore suffices to show that the determinant of $\mathbb{V}$ is 
non-negative. The direct calculation of the determinant yields
\begin{multline}\label{SI 3}
\frac{{{\rm Det}} \; \mathbb{V}}{4}=(1- |S_{LL}|^2)(1-|S_{PP}|^2)\\ - |S_{LP}|^4 -|S_{PL}|^4 + |S_{LP}|^2 |S_{PL}|^2.
\end{multline}
Now, we distinguish two cases. First, we assume $|S_{LP}|^2 \geq |S_{PL}|^2$ and rewrite (\ref{SI 3}) as 
\begin{equation}
\begin{split}
\frac{{{\rm Det}} \;  \mathbb{V}}{4} &  = (|S_{LP}|^2 + |S_{LR}|^2)(|S_{LP}|^2 + |S_{RP}|^2)\\
                                     & \qquad - |S_{LP}|^4 -|S_{PL}|^4 + |S_{LP}|^2 |S_{PL}|^2\\
                                     &  = |S_{LP}|^2 |S_{RP}|^2 + |S_{LR}|^2 |S_{LP}|^2 + |S_{LR}|^2 |S_{RP}|^2\\
                                     & \qquad  + |S_{PL}|^2 (|S_{LP}|^2 - |S_{PL}|^2) \geq 0.
\end{split}
\end{equation}
Here, we used the sum rule
\begin{equation}\label{SI 4}
\sum_{A=L,R,P} |S_{AB}|^2 = \sum_{A=L,R,P} |S_{BA}|^2 =1,
\end{equation}
which is a direct consequence of the unitarity of the scattering matrix $\mathbb{S}$. On the other hand, if $|S_{LP}|^2 < |S_{PL}|^2$, we can use (\ref{SI 4}) to write 
\begin{equation}
\begin{split}
\frac{{{\rm Det}} \; \mathbb{V}}{4} & = (|S_{PL}|^2 + |S_{RL}|^2)(|S_{PL}|^2 + |S_{PR}|^2)\\
                                    & \qquad- |S_{LP}|^4 -|S_{PL}|^4 + |S_{LP}|^2 |S_{PL}|^2\\
                                    &  = |S_{PL}|^2 |S_{PR}|^2 + |S_{RL}|^2 |S_{PL}|^2 + |S_{RL}|^2 |S_{PR}|^2\\
                                    & \qquad  + |S_{LP}|^2 (|S_{PL}|^2 - |S_{LP}|^2) \geq 0.
\end{split}
\end{equation}
In conclusion, we have shown that ${{\rm Det}}\; \mathbb{V} \geq 0$ and therefore the proof that $\mathbb{K}'$ is positive semidefinite is complete.